\documentclass[preprint,floatfix,noshowpacs]{revtex4}%
\usepackage[dvips]{graphicx}
\usepackage{amsmath,amssymb}
\usepackage{amsmath}
\usepackage{amsfonts}
\usepackage{amssymb}%
\begin{document}
\title{The pion distribution amplitude and the pion-photon transition form factor in
a nonlocal chiral quark model}
\author{D. G\'{o}mez Dumm$^{a,b}$, S. Noguera$^{c}$, N. N. Scoccola$^{b,d,e}$ and S.
Scopetta$^{f}$}
\affiliation{$^{a}$ IFLP, CONICET $-$ Dpto.\ de F\'{\i}sica, Universidad
Nacional de La Plata, C.C. 67, (1900) La Plata, Argentina}
\affiliation{$^{b}$ CONICET, Rivadavia 1917, (1033) Buenos Aires,
Argentina}
\affiliation{$^{c}$ Departamento de F\'{\i}sica Te\'orica and Instituto de
F\'{\i}sica Corpuscular, Universidad de Valencia-CSIC, E-46100 Burjassot
(Valencia), Spain}
\affiliation{$^{d}$ Physics Department, Comisi\'on Nacional de
Energ\'{\i}a At\'omica, } \affiliation{Av.\ Libertador 8250, (1429) Buenos
Aires, Argentina}
\affiliation{$^{e}$ Universidad Favaloro, Sol{\'{\i}}s 453, (1078) Buenos
Aires, Argentina}
\affiliation{$^{f}$ Dipartimento di Fisica, Universit\`{a} di Perugia, and
INFN, Sezione di Perugia, via A. Pascoli, I-06100 Perugia, Italy}

\begin{abstract}
We study the pion Distribution Amplitude ($\pi$DA) in the context of a
nonlocal chiral quark model. The corresponding Lagrangian reproduces the
phenomenological values of the pion mass and decay constant, as well as
the momentum dependence of the quark propagator obtained in lattice
calculations. It is found that the obtained $\pi$DA has two symmetric
maxima, which arise from the new contributions generated by the nonlocal
character of the interactions. This $\pi$DA is applied to leading order
and next-to-leading order calculations of the pion-photon transition form
factor. Implications of the results are discussed.

\end{abstract}

\maketitle

\section{Introduction}

The pion Distribution Amplitude ($\pi$DA) is a fundamental theoretical
ingredient in the description of exclusive high-energy processes. The
simplest hard exclusive process determined by the $\pi$DA is the
transition $\pi\rightarrow\gamma\gamma^{\ast }$ at high photon virtuality
$Q^{2}$, since for this process the pion is the only hadron involved; on
the other hand, the large $Q^{2}$ behavior of the related Pion Transition
Form Factor ($\pi$TFF) is well known from perturbative
QCD~\cite{Efremov:1979qk, Lepage:1979zb}. The $\pi$TFF can be measured for
both space-like and time-like momentum transfers through
the processes $e^{+}e^{-}\rightarrow e^{+}e^{-}\pi^{0}$ and $e^{+}%
e^{-}\rightarrow\pi^{0}\gamma$, respectively. The corresponding
experimental status has been improved in the last years, since old results
from CELLO~\cite{Behrend:1990sr} (covering a space-like momentum transfer
region $0.68<Q^{2} <2.17\ \mbox{GeV}^{2}$) and CLEO~\cite{Gronberg:1997fj}
($1.64<Q^{2} <7.9\ \mbox{GeV}^{2}$) have now been complemented with data
from the BABAR~\cite{:2009mc} and BELLE~\cite{Uehara:2012ag}
Collaborations, which cover pion virtualities ranging from $4$ to
35~GeV$^{2}$. While the old data suggested that the $\pi$TFF reaches its
asymptotic behavior for $Q^{2}$ values of the order of a few GeV$^{2}$,
the new BABAR data exhibit a steeper growth, indicating that the
asymptotic QCD limit is crossed at $Q^{2}\sim10$~GeV$^{2}$. The BELLE data
show instead a slower growth, in which the $\pi$TFF seems to cross the
asymptotic limit at $Q^{2}\sim20$~GeV$^{2}$. In addition, the BABAR
Collaboration has recently measured the $\eta$ and $\eta^{\prime}$
transition form factors~\cite{:2011hk}; the data show in this case a mild
behavior, approaching from below the corresponding asymptotic QCD limit
for large $Q^{2}$ values. In any case, owing to the relatively large
errors, it could be said that present experimental data are compatible
with each other, and still more accurate measurements would be needed in
order to firmly establish the behavior of the $\pi$TFF in the region of
intermediate and large $Q^{2}$.

In view of the new experimental results, a significant theoretical effort
has been carried out towards the obtention of theoretical predictions for
the $\pi$DA and $\pi$TFF. First analyses have proposed a flat $\pi$DA,
i.e.~$\phi_{\pi}\left(  x\right)  =1$~\cite{Radyushkin:2009zg,
Polyakov:2009je}. This scenario is compatible with QCD sum
rules~\cite{Chernyak:1981zz} and lattice QCD
results~\cite{DelDebbio:2005bg,Braun:2006dg}, which lead to values for the
second moment of the $\pi$DA that are large in comparison with that
obtained using the asymptotic $\pi$DA $\phi_{\pi}\left(  x\right)
=6x(1-x)$. A constant $\pi$DA is in fact obtained within effective
theories such as the Nambu$-$Jona-Lasinio (NJL)
model~\cite{RuizArriola:2002bp,Courtoy:2007vy,Courtoy:2010qn} and the
\textquotedblleft spectral\textquotedblright\ quark
model~\cite{RuizArriola:2003bs}. A formalism which connects the
experimental parametrization of the $\pi$TFF at low photon virtuality with
the description of the $\pi$TFF at high photon virtuality using a flat
$\pi$DA has been developed in Ref.~\cite{Noguera:2010fe}. Within this
formalism, a good agreement with the experimental pattern is achieved
after the inclusion of a correction carrying an extra power of $1/Q^{2}$,
which is needed in order to reproduce the data in the region
$1<Q^{2}<15$~GeV$^{2}$. With the same ingredients, in the context of the
NJL model a good description of the $\eta $TFF can be
obtained~\cite{Noguera:2011fv}. Finally, other analyses carried out within
quark models can be found in Refs.~\cite{Praszalowicz:2001wy,
Dorokhov:2013xpa}. In all quark model approaches the $\pi$DA is provided
by the models, whose parameters are fitted from other physical quantities.

The $\pi$DA and the $\pi$TFF have also been studied within the Non Local
Condensates Sum Rule (NLC-SR) and Light Cone Sum Rule (LCSR)
approaches~\cite{Bakulev:2001pa,
Mikhailov:2009kf,Mikhailov:2010ud,Bakulev:2011rp,Agaev:2010aq,Agaev:2012tm}.
These calculations use similar ingredients, introducing corrections with
extra powers of $1/Q^{2}$ in order to describe the data through the twist
4 and 6 contributions. While in Refs.~\cite{Agaev:2010aq,Agaev:2012tm} a
good description of the experimental results is obtained, in
Refs.~\cite{Bakulev:2011rp} it is claimed that in order to reproduce the
data from BABAR one would need some enhancement mechanism that cannot be
explained within the standard QCD scheme based on collinear factorization.
A study of the $\pi$DA is also presented in Refs.~\cite{Kroll:2010bf,
Huang:2013yya}, starting from the pion leading twist wave function. In
general, in all these works the $\pi$DA is parametrized in terms of an
expansion in a series of Gegenbauer polynomials. This expansion is
truncated keeping the first few polynomials, and the corresponding
coefficients are treated as parameters to be adjusted.

The aim of this work is to study the $\pi$DA and the $\pi$TFF within the
framework of a nonlocal Nambu$-$Jona-Lasinio model (nlNJL). The NJL model
is a simple scheme based on the QCD feature of dynamical chiral symmetry
breaking, in which quarks interact through a local, chiral invariant
four-fermion coupling. The local nature of this interaction allows to
obtain simple solutions of the corresponding Dyson-Schwinger and
Bethe-Salpeter equations. However, the main drawbacks of the model are
direct consequences of the locality: a definite prescription is needed in
order to regularize ultraviolet loop divergences, and the model is
nonconfining. The nlNJL model represents an improvement over the local
theory. Indeed, it can be seen that nonlocal form factors regularize the
model in such a way that anomalies are preserved and charges are properly
quantized, and there is no need to introduce extra cut-offs. In fact,
nonlocality arises naturally in quantum field theory when the interactions
involve large coupling constants.

The starting point in our analysis will be a Lagrangian theory that
includes couplings between nonlocal quark currents. In this way, our
formalism ensures the preservation of fundamental symmetries (chiral,
Poincar\'{e} and local electromagnetic gauge invariances) that guarantee
the proper normalization of the $\pi$DA. In the framework of a Lagrangian
theory, the three main ingredients of a nonperturbative analysis that
involves photons and the pion are: $i)$ the quark propagator, which obeys
the Dyson-Schwinger equation; $ii)$ the description of the pion as a bound
state of a Bethe-Salpeter equation (BSE); $iii)$ a prescription for
introducing the electroweak interaction that preserves gauge symmetry.
Owing to the chiral symmetry, the kernels of the equations appearing in
$i)$ and $ii)$ are not independent~\cite{Delbourgo:1979me}. The
Dyson-Schwinger equation leads to momentum dependences in the quark
propagators through its mass and its wave function renormalization. In our
scheme the gluons have been integrated out (we have only flavor
interaction between quarks), and confinement is obtained from the
structure of the quark propagator and by limiting the Fock space to color
singlet states. The pion is described in a consistent way by solving the
BSE, and it shows up as a Goldstone boson associated with the spontaneous
breakdown of the chiral symmetry. Finally, the couplings involving photons
and weak bosons are implemented by imposing local gauge invariance in the
Lagrangian. Therefore, we must gauge not only the kinetic term, but also
the nonlocal quark currents in the interaction terms.

The quark propagator is taken as one of the main ingredients of our model.
The reason is that one has direct information on this propagator from the
fundamental QCD theory, since the momentum dependences of quark mass and
wave function renormalization have been calculated in lattice
QCD~\cite{Parappilly:2005ei,Bowman2003}. Our Lagrangian is in fact the
minimal framework that allows to incorporate the full momentum dependence
obtained through these lattice calculations. In this way, our model can be
seen as an extension of nonlocal NJL models analyzed in previous
works~\cite{Bowler:1994ir,Scarpettini:2003fj,GomezDumm:2006vz,Golli:1998rf,
Rezaeian:2004nf,Praszalowicz:2001wy}, but with a particular philosophy.
The model considered here has been proposed in Ref.~\cite{Noguera:2005ej},
and then it has been successfully applied to the analysis of different
hadronic observables~\cite{Noguera:2005cc,Noguera:2008,Dumm:2010hh}.

Once the Lagrangian theory has been defined, it is possible to obtain the
$\pi$DA from a fundamental calculation. The main difficulty to be
solved is that the bilocal axial current present in the definition of the
$\pi$DA will be dressed by the nonlocal interaction. To deal with this
problem we rely on the basic physical idea beyond the factorization of the
$\pi$TFF into hard and soft contributions for high $Q^{2}\,$: the struck
quark loses its high momentum before being able to interact with the
remaining quarks and gluons of the hadron. This situation will be
implemented here by considering the bilocal current associated to the
$\pi$DA as a current coupled to an external fictitious probe carrying the
adequate quantum numbers.

The $\pi$DA provides the dominant twist two contribution to the $\pi$TFF.
Corrections to this term will be introduced considering contributions that
carry extra powers of $1/Q^{2}$ (we will include $1/Q^{4}$ and $1/Q^{6}$
terms). Therefore, in our scheme we have a fixed $\pi$DA and two free
parameters in the $\pi$TFF. This is in contrast with the program followed
in Refs.~\cite{Mikhailov:2009kf, Mikhailov:2010ud, Bakulev:2011rp,
Agaev:2010aq,Agaev:2012tm,Huang:2013yya,Kroll:2010bf}, where the $\pi$DA
is parametrized in terms of a expansion in Gegenbauer polynomials with
free coefficients and the twist four and six corrections are constrained
by sum rule techniques.

The present paper is organized as follows. In section~\ref{Sec_formalism}
we describe the connection between the $\pi$TFF and the $\pi$DA, we
present the model Lagrangian and we quote our analytical results for the
$\pi$DA. In section~\ref{Sec_DA} we show and discuss the numerical results
for the $\pi$DA obtained within our model. The dependence on the
transverse momentum $k_{T}$ and the connection with the light cone wave
functions are discussed in section~\ref{Sec_kT}. In section~\ref{Sec_TFF}
the results obtained for the $\pi$TFF are analyzed. Finally, in
section~\ref{Sec_Conclusion} we sketch our conclusions. Details of the
calculations, including some relevant analytical expressions, can be found
in Appendices \ref{App_NL_NJL} and \ref{App_QCD_Evolution}.

\section{Formalism}

\label{Sec_formalism}

\subsection{Generalities on the evaluation of the $\pi$TFF and $\pi$DA in
effective quark models}

\begin{figure}[ptb]
\begin{center}
\includegraphics[
width=3.5in]{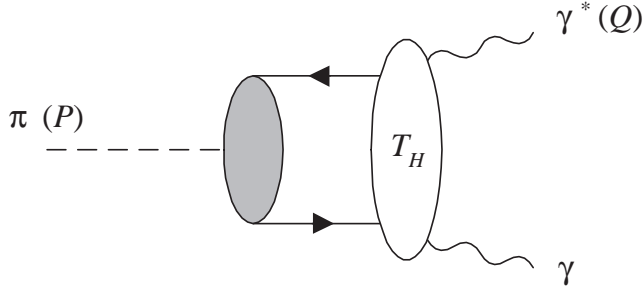}
\caption{Schematic structure of the QCD factorization for the $\pi
\rightarrow\gamma\,\gamma^{\ast}$ process.}
\label{Fig_Pi_Gam_GamSt}
\end{center}
\end{figure}

At stated, the transition form factor for the process $\pi^{0}\rightarrow
\gamma\gamma^{\ast}$ at large photon virtuality is basically determined by the
pion distribution amplitude. At the leading order in powers of $1/Q^{2}$ one
has
\begin{equation}
F(Q^{2})=\frac{\sqrt{2}f_{\pi}}{3\,Q^{2}}\int_{0}^{1}dx\,T_{H}(x,Q^{2}%
,\mu)\,\phi_{\pi}(x,\mu)\ , \label{PionTFF.01}%
\end{equation}
where $f_{\pi}=0.131$ GeV. Here the function $T_{H}(x,Q^{2},\mu)$, which
includes both photon vertices (see Fig.~\ref{Fig_Pi_Gam_GamSt}), accounts for
the hard contributions to the process and can be calculated from perturbative
QCD. In the modified minimal subtraction ($\overline{\text{MS}}$) scheme, up
to the NLO in the strong coupling, one obtains~\cite{delAguila:1981nk,
Braaten:1982yp}
\begin{equation}
T_{H}^{\text{NLO}}\left(  x,Q^{2},\mu\right)  =\frac{1}{x}\left\{
1+C_{F}\frac{\alpha_{\mathrm{s}}\left(  \mu\right)  }{4\,\pi}\left[  \ln
^{2}x-\frac{x\,\ln x}{1-x}-9+\left(  3+2\ln x\right)  \ln\frac{Q^{2}}{\mu^{2}%
}\right]  \right\}  \ , \label{PionTFF.03}%
\end{equation}
with $C_F = 4/3$ for $N_c=3$. On the other hand, $\phi_{\pi}(x,\mu)$
stands for the $\pi$DA, which involves the soft, nonperturbative
contributions to the form factor [in Eq.~(\ref{PionTFF.03}), the symmetry
property $\phi_{\pi}(1-x,\mu)=\phi_{\pi}(x,\mu)$ has been used]. One can
take this distribution amplitude from some theoretical model for the pion
or, alternatively, it can be parametrized phenomenologically. Finally, the
parameter $\mu$ is the renormalization and factorization scale, which will
be set here by $\mu^2 = Q^2$. Different relations between $\mu^2$ and
$Q^2$ have been considered in Ref.~\cite{Agaev:2010aq}. In fact, our
results do not show a significant numerical variation for these different
choices. For simplicity, in the following we will omit the $\mu$
dependence in $\phi_{\pi}(x,\mu)$ unless necessary.

We will postpone the analysis of the $\pi$TFF to Sec.~V, and concentrate
now in the evaluation of the $\pi$DA within the framework of an effective
quark model. By definition, the $\pi$DA $\phi(x)$ is given by
\begin{equation}
i\sqrt{2}\,f_{\pi}~\phi_{\pi}\left(  x\right)  \ =\ \int\frac{dz^{-}}{2\pi
}~e^{iP^{+}z^{-}\left(  x-\frac{1}{2}\right)  }~\left.  \left\langle
0\right\vert \bar{\psi}\left(  -\frac{z}{2}\right)  \gamma^{+}\gamma_{5}%
\tau^{-}\psi\left(  \frac{z}{2}\right)  \left\vert \pi^{-}\left(  P\right)
\right\rangle \right\vert _{z^{+}=0,\,\vec{z}_{T}=0}\ , \label{PionDA.01}%
\end{equation}
where we have introduced the light front components $P^{+}$, $z^{-}$,
$\gamma^{+}$, choosing a frame in which $\vec{P}_{T}=0$. For any
four-vector $a^{\mu},$ the light front components are defined by
$a^{\pm}=(a^{0}\pm a^{3})/\sqrt{2}$, while $\vec{a}_{T}\equiv
(a^{1},a^{2})$. As it is well known, $x$ becomes the fraction of the $+$
component of the momentum carried by the struck quark in the meson, and
its support is the interval $\left[  0,1\right]  $.

Recalling that the pion decay constant can be defined by
\begin{equation}
f_{\pi}=\frac{1}{i\sqrt{2}\,P^{+}}~\left\langle 0\right\vert \bar{\psi}\left(
0\right)  \gamma^{+}\gamma_{5}\tau^{-}\psi\left(  0\right)  \left\vert \pi
^{-}\left(  P\right)  \right\rangle \ , \label{PionDA.03}%
\end{equation}
from Eq.~(\ref{PionDA.01}) one obtains for $\phi_{\pi}\left(  x\right)  $ the
sum rule
\begin{equation}
\int_{0}^{1}dx\,\phi_{\pi}\left(  x\right)  =1\ . \label{PionDA.05}%
\end{equation}
It is worth stressing that this is not a normalization condition to be
imposed, but a result that has to be fulfilled by any well defined model.

\subsection{$\pi$DA in a nonlocal NJL model with wavefunction renormalization}

We consider here a nonlocal covariant SU(2) chiral quark model that includes
wave function renormalization in the quark propagator. The corresponding
Euclidean action reads~\cite{Noguera:2005ej,Noguera:2008}
\begin{equation}
S_{E}=\int d^{4}y\ \left\{  \bar{\psi}(y)\left(  -i\rlap/\partial
+m_{c}\right)  \psi(y)-\frac{G_{S}}{2}\Big[j_{a}(y)j_{a}(y)+j_{P}%
(y)j_{P}(y)\Big]\right\}  \ . \label{action}%
\end{equation}
Here $m_{c}$ is the current quark mass, which is assumed to be the same for
$u$ and $d$ quarks, while the nonlocal currents $j_{a}(y)$, $j_{P}(y)$ are
given by
\begin{align}
j_{a}(y)  &  =\int d^{4}z\ \mathcal{G}(z)\ \bar{\psi}\left(  y+\frac{z}%
{2}\right)  \ \Gamma_{a}\ \psi\left(  y-\frac{z}{2}\right)  \ ,\nonumber\\
j_{P}(y)  &  =\int d^{4}z\ \mathcal{F}(z)\ \bar{\psi}\left(  y+\frac{z}%
{2}\right)  \ \frac{i{\overleftrightarrow{\rlap/\partial}}}{2\ \varkappa_{p}%
}\ \psi\left(  y-\frac{z}{2}\right)  \ , \label{cuOGE}%
\end{align}
where $\Gamma_{a}=(\leavevmode\hbox{\small1\kern-3.8pt\normalsize1},i\gamma
_{5}\vec{\tau})$ and $u(y^{\prime}){\overleftrightarrow{\partial}%
}v(y)=u(y^{\prime})\partial_{y}v(y)-\partial_{y^{\prime}}u(y^{\prime})v(y)$.
The nonlocal character of the interactions is provided by the covariant
vertex form factors $\mathcal{G}(z)$ and $\mathcal{F}(z)$ in
Eq.~(\ref{cuOGE}). In the mean field approximation these functions
determine the momentum dependence of the mass and wave function
renormalization in the quark propagator,
\begin{equation}
\mathcal{D}_{0}(p)^{-1}\ =\ \frac{z_{p}}{-\not \!  \!p\,+m_{p}}\ ,
\label{qprop}%
\end{equation}
where
\begin{equation}
z_{p}\equiv z\left(  p\right)  =\left(  1-\bar{\sigma}_{2}\ f_{p}\right)
^{-1}\ ,\qquad m_{p}\equiv m\left(  p\right)  =z_{p}\left(  m_{c}+\bar{\sigma
}_{1}\ g_{p}\right)  \ . \label{zm}%
\end{equation}
The functions $g_{p}$ and $f_{p}$ in these equations are the Fourier
transforms of $\mathcal{G}(z)$ and $\mathcal{F}(z)$, while $\bar{\sigma}%
_{1,2}$ are the mean field values of the scalar fields associated with the
currents $j_{0}(y)$ and $j_{P}(y)$, respectively. The main point here is that
starting from a given expression for $m_{p}$ and $z_{p},$ based in our case on
lattice results, we can use Eq. (\ref{zm}) for extracting the related $g_{p}$
and $f_{p}$ functions. The mean field values $\bar{\sigma}_{1,2}$ are related
to the values of $m_{p}$ and $z_{p}$ at $p=0$ through
\begin{equation}
\bar{\sigma}_{2}=1-\frac{1}{z\left(  0\right)  }\ ,\qquad\qquad\bar{\sigma
}_{1}=\frac{m\left(  0\right)  }{z\left(  0\right)  }-m_{c} \ .
\end{equation}
Following Ref.~\cite{Noguera:2008}, we choose $m_{p}$ and $z_{p}$ as%
\begin{align}
m_{p}  &  =m_{c}+\frac{\alpha_{m}}{1+\left(  p^{2}/\Lambda_{0}^{2}\right)
^{3/2}}\ ,\nonumber\\
z_{p}  &  =1+\frac{\alpha_{z}}{\left[  1+\left(  p^{2}/\Lambda_{1}^{2}\right)
\right]  ^{5/2}}\ , \label{parametrization_set2}%
\end{align}
where $m_{c}=2.37$ MeV, $\alpha_{m}=309$ MeV, $\alpha_{z}=-0.3$, $\Lambda
_{0}=850$ MeV and $\Lambda_{1}=1400$~MeV. This parametrization allows to
reproduce very well the momentum dependence of the quark propagator mass and
wave function renormalization obtained in lattice calculations
\cite{Parappilly:2005ei,Bowman2003}, providing at the same time the proper
physical values for the pion mass and decay constant~\cite{Noguera:2008}.

Given this effective model for the strong interactions at low energies, one
can explicitly evaluate the $\pi$DA from Eq.~(\ref{PionDA.01}). Since the
amplitude involves a bilocal axial vector current, one should introduce into
the effective action in Eq.~(\ref{action}) a coupling to an external axial
gauge field $a_{\mu}$. For a local theory this can be done by performing the
replacement%
\begin{equation}
\partial_{\mu}\rightarrow\partial_{\mu}+i\ \mathcal{A}_{\mu}(y)\ ,
\end{equation}
where, according to the quantum numbers of the $\pi^{-}$ field,
\begin{equation}
\mathcal{A}_{\mu}(y)=\tau^{-}\gamma_{5}\,a_{\mu}(y)\ .
\end{equation}
In the case of the above described nlNJL model the situation is more
complicated since the inclusion of gauge interactions implies a change not
only in the kinetic terms in the Lagrangian but also in the nonlocal currents
appearing in the interaction terms. If $y$ and $z$ denote the space variables
in the definitions of the nonlocal currents [see Eq.~(\ref{cuOGE})], one has
\begin{align}
\psi(y-z/2)\  &  \rightarrow W\left(  y,y-z/2\right)  \ \psi
(y-z/2)\ ,\nonumber\\
\psi^{\dagger}(y+z/2)\  &  \rightarrow\psi^{\dagger}(y+z/2)\ W\left(
y+z/2,y\right)  \ . \label{gauge}%
\end{align}
Here the function $W(s,t)$ is defined by
\begin{equation}
W(s,t)\ =\ \mathrm{P}\;\exp\left[  i\ \int_{s}^{t}dr_{\mu}\ \mathcal{A}_{\mu
}(r)\right]  \ , \label{intpath}%
\end{equation}
where $r$ runs over an arbitrary path connecting $s$ with $t$.

This procedure has been analyzed in detail within nlNJL models, in
particular regarding the calculation of the pion decay
constant~\cite{Bowler:1994ir,Scarpettini:2003fj,Noguera:2005ej}, see
Eq.~(\ref{PionDA.03}). The situation is similar for the case of the
bilocal axial current in the definition of the $\pi$DA. In fact, the basic
physical idea beyond the factorization of the $\pi$TFF into hard and soft
contributions is that for high $Q^{2}$ the struck quark loses its high
momentum before being able to interact with the remaining quarks and
gluons of the hadron ($Q^{2}\sim1$ GeV$^{2}$ implies a time scale of the
order of $10^{-24}$ s). Therefore, the nonlocal interaction does not see
the struck quark but only the quarks in the hadron before and after the
photon absorption-emission process. This can be effectively implemented by
introducing an external fictitious probe carrying the adequate quantum
numbers, which in our case is an axial gauge field (a similar situation
has been studied in the case of the pion Parton Distribution, see
Refs.~\cite{Noguera:2005ej,Noguera:2005cc}). Thus, the axial vertex in
Eq.~(\ref{PionDA.01}) will become dressed by the nonlocal interaction,
irrespective of whether the quark current is a local or a bilocal one (as
in this case).

The steps to be followed in the explicit calculation of the $\pi$DA within the
nlNJL model are detailed in Appendix~\ref{App_NL_NJL}. We quote here the
resulting expression
\begin{equation}
\phi_{\pi}\left(  x\right)  =\frac{2\sqrt{2}\,N_{c}\,g_{\pi q\bar q}}{f_{\pi}}%
\int\frac{dw\,d^{2}k_{T}}{\left(  2\pi\right)  ^{4}}\ F\left(  w,x,k_{T}%
\right)  \ , \label{DA.01}%
\end{equation}
where $g_{\pi q\bar q}$ stands for an effective quark-meson coupling
constant (see Appendix~\ref{App_NL_NJL}). It is convenient to separate the
integrand in Eq.~(\ref{DA.01}) into two pieces,
\begin{equation}
F\left(  w,x,k_{T}\right)  \ =\ F_{1}\left(  w,x,k_{T}\right)  +F_{2}\left(
w,x,k_{T}\right)  \ . \label{terms}%
\end{equation}
The explicit expressions for these functions are
\begin{eqnarray}
F_{1}\left(  w,x,k_{T}\right)   &  = & \frac{g_{k}}{2}\frac{z_{k_{+}}\,z_{k_{-}}%
}{D_{k_{+}}D_{k_{-}}}\left(  \frac{1}{z_{k_{+}}}+\frac{1}{z_{k_{-}}}\right)
\left[  \left(  1-x\right)  \,m_{k_{+}}+x\,m_{k_{-}}\right]  \ ,
\label{DA.10}\\[3mm]
F_{2}\left(  w,x,k_{T}\right)   &  = &
g_{k}\frac{z_{k_{+}}z_{k_{-}}}{D_{k_{+}%
}D_{k_{-}}}\Big\{[k_{+}\cdot k_{-}+m_{k_{+}}\,m_{k_{-}}]\,\nu_{1}\; - \nonumber\\
& &  \qquad\qquad\qquad k\cdot\left[
k_{+}\,m_{k_{-}}-k_{-}\,m_{k_{+}}\right]  \,\nu
_{2}\Big\}-\frac{m_{k}\,z_{k}}{D_{k}\,\bar{\sigma}_{1}}\ \nu_{1}\ ,
\label{DA.11}%
\end{eqnarray}
where we have defined $k_{\pm}=k\pm P/2$ and $D_{k}=k^{2}+m_{k}^{2}$. In
terms of the variables $w$ and $k_{T}$ we have
\[
k^{2}=-i\,w\left(  x-\frac{1}{2}\right)  +m_{\pi}^{2}\left(  x-\frac{1}%
{2}\right)  ^{2}+k_{T}^{2}\ ,\qquad k\cdot P=-i\,\frac{w}{2}\ .
\]
Finally, the functions $\nu_{1}$ and $\nu_{2}$ in Eq.~(\ref{DA.11}) are
given by
\begin{align}
\nu_{1}  &  =\frac{\left(  x-\frac{1}{2}\right)  }{k\cdot P}\Big[\frac
{m_{k_{+}}}{z_{k_{+}}}+\frac{m_{k_{-}}}{z_{k_{-}}}-2\frac{m_{k}}{z_{k}}%
+m_{\pi}^{2}\,\bar{\sigma}_{1}\,\alpha_{g}^{-}\Big]+\bar{\sigma}_{1}%
\,\alpha_{g}^{-}\ ,\nonumber\\
\nu_{2}  &  =\frac{\left(  x-\frac{1}{2}\right)  }{k\cdot P}\Big[\frac
{1}{z_{k_{-}}}-\frac{1}{z_{k_{+}}}+m_{\pi}^{2}\,\bar{\sigma}_{2}\,\alpha
_{f}^{+}\Big]+\bar{\sigma}_{2}\,\alpha_{f}^{+}\ , \label{nu12}%
\end{align}
where $\alpha_{g}^{-}$ and $\alpha_{f}^{+}$ depend in general on the
integration path in Eq.~(\ref{intpath}). If one takes a straight line path
the corresponding explicit expressions read
\[
\alpha_{g}^{-}=\int_{0}^{1}d\lambda\ \frac{\lambda}{2}\ g_{k-\lambda
P/2}^{\,\prime}-\int_{-1}^{0}d\lambda\ \frac{\lambda}{2}\ g_{k-\lambda
P/2}^{\,\prime}\ ,\qquad\alpha_{f}^{+}=\int_{-1}^{1}\ d\lambda\ \frac{\lambda
}{2}\ f_{k-\lambda P/2}^{\,\prime}\ .
\]

\subsection{LO and NLO evolution of the $\pi$DA}

Once the $\pi$DA $\phi(x)$ is known at a given $\mu_{0}$ scale, its
evolution up to a new scale $\mu$ can be obtained from perturbative
QCD~\cite{Efremov:1979qk,Lepage:1979zb}. In order to calculate this
evolution (we denote now explicitly the $\mu$ dependence of the $\pi$DA),
it is convenient to expand $\phi_{\pi}(x,\mu)$ in a series of Gegenbauer
polynomials,
\begin{equation}
\phi_{\pi}(x,\mu)=6x(1-x)\sum_{n=0}^{\infty}a_{n}(\mu)\,C_{n}^{3/2}(2x-1)\ .
\label{DA.31}%
\end{equation}
{}From the orthogonality relations satisfied by these polynomials one gets the
coefficients at the $\mu_{0}$ scale, namely
\begin{equation}
a_{n}\left(  \mu_{0}\right)  =\frac{2\left(  2n+3\right)  }{3\left(
n+1\right)  \left(  n+2\right)  }\int_{0}^{1}dx\,C_{n}^{3/2}(2x-1)\,\phi_{\pi
}\left(  x,\mu_{0}\right)  \ . \label{DA.33}%
\end{equation}
If $\phi_{\pi}(x,\mu_{0})$ satisfies the sum rule Eq.~(\ref{PionDA.05}),
then the first coefficient $a_{0}(\mu_{0})$ has to be equal to $1$. Thus
all the information from the pion effective model is included in the
remaining coefficients $a_{n}(\mu_{0})$, with $n=2,4,\ldots\ $. At the LO
in the strong coupling $\alpha_{s}$ the coefficients turn out to be
renormalized multiplicatively,
\begin{equation}
a_{n}^{\text{LO}}(\mu)=a_{n}(\mu_{0})\,E_{n}^{\text{LO}}(\mu,\mu_{0})\ ,
\label{DA.34}%
\end{equation}
whereas at the NLO the evolution equations for different coefficients get
mixed, and the pattern becomes more complicated. One
has~\cite{Agaev:2010aq}
\begin{align}
a_{n}^{\text{NLO}}(\mu) = a_{n}(\mu_{0})\,E_{n}^{\text{NLO}}(\mu,\mu_{0}) +
\frac{\alpha_{s}(\mu)}{4\pi}\sum_{k=0}^{n-2}a_{k} (\mu_{0})\,E_{k}^{\text{LO}%
}(\mu,\mu_{0})\,d_{n}^{k}(\mu,\mu_{0})\ . \label{DA.35}%
\end{align}
Explicit expressions for the renormalization factors $E_{n}^{\text{LO}}%
(\mu,\mu_{0})$, $E_{n}^{\text{NLO}}(\mu,\mu_{0})$, as well as for the
off-diagonal mixing coefficients $d_{n}^{k}(\mu,\mu_{0})$ in the
$\overline{\text{MS}}$ scheme, are collected in
Appendix~\ref{App_QCD_Evolution}. Usually the calculation of a few
coefficients $a_{n}(\mu)$ is sufficient to get a good estimate of the $\pi$DA
at the scale $\mu$ using Eq.~(\ref{DA.31}).

\section{Pion Distribution Amplitude}

\label{Sec_DA}

Our result for the $\pi$DA, Eq.~(\ref{DA.01}), is plotted in
Fig.~\ref{Fig_DA_model} (solid line), where the contributions coming from
Eqs.~(\ref{DA.10}) and (\ref{DA.11}) are also separately shown (dashed and
dotted lines, respectively). One observes that the full result has two
symmetric maxima. This feature is also shown by the $\pi$DA calculated in
Refs.~\cite{Chernyak:1981zz,Bakulev:2001pa}, but in our case the two
maxima are much closer to $x=0.5$. From the curves it is seen that this
shape arises from the term in Eq.~(\ref{DA.11}), which is a genuine
nonlocal contribution.

Our calculation is performed in Euclidean space. To check the consistency
of our scheme we rely on the verification of the following fundamental
properties of the $\pi$DA: (i) the $\pi$DA has to be invariant under the
exchange $x\leftrightarrow(1-x)$; (ii) the $\pi$DA has support in the
interval $[0,1]$; (iii) the sum rule Eq.~(\ref{PionDA.05}) has to be
fulfilled.

The first property is a consequence of isospin symmetry. It can be easily
checked from the analytical expressions in Eqs.~(\ref{DA.01}-\ref{nu12}).

Concerning the second property, we notice that it can be associated to the
Wick rotation in cases where an exact solution can be
obtained~\cite{Noguera:2010fe}. Let us assume that quark masses do not
depend on the momentum, and let us write the denominators $D_{k^{\pm}}$ in
Eqs.~(\ref{DA.10}-\ref{DA.11}) in Minkowski space:
\begin{subequations}
\label{02.23}%
\begin{align}
D_{k^{-}} = \,\left(  k_{-}^{2}-m^{2}+i\epsilon\right)   &  =-2P^{+}\left(
1-x\right)  \left[  \left(  k^{-}-\frac{P^{-}}{2}\right)  +\frac{\vec
{k}_T^{2}+m^{2}}{2P^{+}\left(  1-x\right)  }-\frac{i\epsilon}%
{2P^{+}\left(  1-x\right)  }\right]  ~,\label{02.23a}\\
D_{k^{+}} = \, \left(  k_{+}^{2}-m^{2}+i\epsilon\right)   &  =2P^{+}x\left[
\left(  k^{-}+\frac{P^{-}}{2}\right)  -\frac{\vec{k}_T^{2}+m^{2}%
}{2P^{+}x}+\frac{i\epsilon}{2P^{+}x}\right]  ~. \label{02.23b}%
\end{align}
We observe that the integration of the function in Eq.~(\ref{DA.10}) with
respect to $k^{-}$ is different from zero only if $0<x<1$. Indeed, we can
perform the Wick rotation in the region $0<x<1$, where it is well defined
according to the positions of the poles determined by Eqs.~(\ref{02.23}),
whereas for $x<0$ and $x>1$ the $\pi$DA will trivially vanish. For a
calculation performed in Euclidean space (as in our case), the poles lie
outside the region of integration, and the loop integrals are formally well
defined. However, for $x<0$ or $x>1$ the positions of the poles do not allow
us to perform the Wick rotation, thus in these regions the result cannot be
connected with the definition of the $\pi$DA in Minkowski space. Consequently,
the integral in Eq.~(\ref{DA.01}) will have physical meaning only for
$x\in[0,1]$.

The last, third property becomes the main consistency check for a
calculation in Euclidean space. Indeed, the fact that the sum rule is
fulfilled when $\phi_{\pi}(x)$ is integrated from 0 to 1 confirms that our
$\pi$DA has the proper support.

\begin{figure}
[ptb]
\begin{center}
\includegraphics[
height=2.7256in,
width=3.8696in
]%
{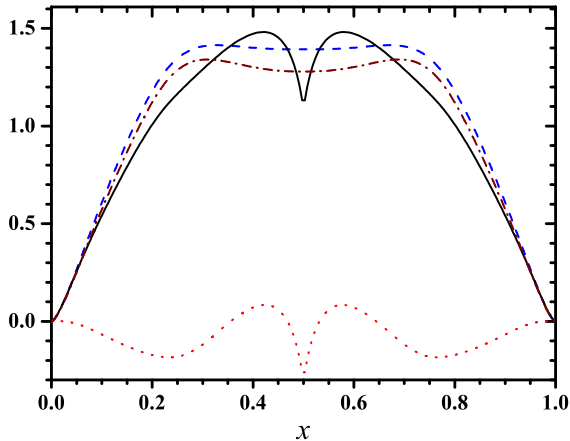}%
\caption{Pion distribution amplitude. The solid line stands for the
$\pi$DA obtained in the present approach, Eq. (\ref{DA.01}), while dashed
and dotted lines correspond to the contributions given by
Eqs.~(\ref{DA.10}) and (\ref{DA.11}), respectively. The dashed-dotted
curve stands for the distribution amplitude defined by Eq.~(\ref{DA.17}),
normalized in order to satisfy the sum rule Eq.~(\ref{PionDA.05}).}%
\label{Fig_DA_model}%
\end{center}
\end{figure}

Within the framework of nlNJL models, several authors have used in the
definition of the $\pi$DA the operator $\gamma^{+}\gamma_{5}$ without
dressing. In our scheme, this choice would correspond to the following
expression for the $\pi$DA:
\end{subequations}
\begin{equation}
\phi_{\pi}^{\left(  0\right)  }\left(  x\right)
=\frac{2\sqrt{2}\,N_c\,g_{\pi
q\bar q}}{f_{\pi}}\int\frac{dw\,d^{2}k_T}{\left(  2\pi\right)  ^{4}}%
\frac{g_{k}\,z_{k_{+}}\,z_{k_{-}}\,\left[  \left(  1-x\right)  \,m_{k_{+}%
}+x\,m_{k_{-}}\right]  }{\left(  k_{+}^{2}+m_{k_{+}}^{2}\right)  \,\left(
k_{-}^{2}+m_{k_{-}}^{2}\right)  }\ . \label{DA.17}%
\end{equation}
It can be seen that in this case the sum rule Eq.~(\ref{PionDA.05}) is not
satisfied. Indeed, in our approach, the usage of Eq.~(\ref{DA.17}) to
evaluate the sum rule yields 0.845 instead of 1. The distribution
amplitude given by Eq.~(\ref{DA.17}), properly normalized to satisfy the
sum rule [i.e.~$\phi _{\pi}^{(0)}(x)/0.845$], is also shown in
Fig.~\ref{Fig_DA_model} (dashed-dotted line). We observe that, except for
a soft depression in the central part, this result is close to the
contribution given by Eq.~(\ref{DA.10}).

Let us consider now the QCD evolution of the $\pi$DA. A crucial point is
the choice of the scale $\mu_{0}$ to be associated to the result provided
by the quark model. In our case this value is fixed by that of the lattice
calculation used to model the quark propagator. According to
Ref.~\cite{Parappilly:2005ei}, we have to take $\mu_{0}=3$~GeV, which is a
large value compared to the scale $\mu_{0}\sim 1$~GeV usually adopted in
model calculations.

\begin{figure}
[ptb]
\begin{center}
\includegraphics[
height=2.8094in,
width=4.0133in
]%
{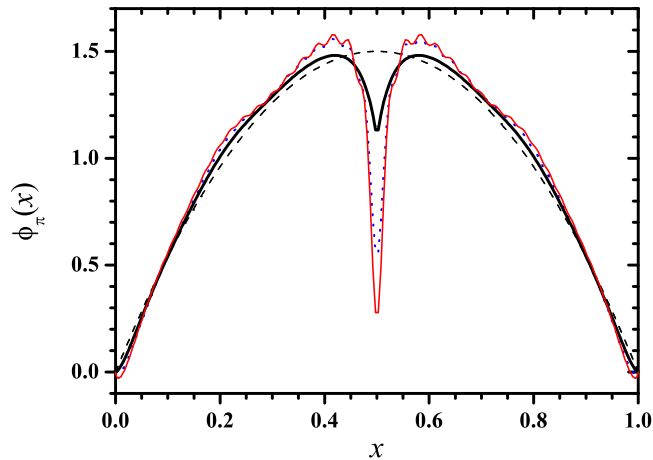}%
\caption{$\pi$DA within our model at $\mu=3$~GeV (thick solid line), and
evolved $\pi$DA at $\mu=1$~GeV, at both LO (dotted line) and NLO (thin
solid line). The dashed line corresponds to the asymptotic $\pi$DA limit.}
\label{Fig_DA_evol}%
\end{center}
\end{figure}

In Fig.~\ref{Fig_DA_evol} we show the distribution amplitude obtained in our
model together with its evolution up to $\mu=1%
\operatorname{GeV}%
$, at LO and NLO. It is seen that the $\pi$DA at $\mu_{0}=3$~GeV is not far
from the asymptotic limit $\phi_{\pi}\left(  x\right)  =6x(1-x)$.
The most significant difference between the results after the LO and NLO
evolutions of the $\pi$DA is that the central minimum decreases significantly;
nevertheless, the two maxima do not separate appreciably. As it is expected,
the $\pi$DA moves away from the asymptotic limit. Another important feature of
the obtained $\pi$DA is that it goes to zero rather fast for $x=0$ and $x=1$,
supporting the idea of suppression of the kinematic endpoints
\cite{Mikhailov:2009kf, Mikhailov:2010ud}. Moreover, this feature is stressed
in the evolution towards smaller values of $\mu$, as it should be expected,
because the predicted $\pi$DA lies below the asymptotic one in this region.

In Table~\ref{Table_a_1} (\ref{Table_a_2}) we quote the first coefficients
of the Gegenbauer expansion obtained with our $\pi$DA at LO (NLO), while
in Table~\ref{Table_a_3} the values obtained by other authors at $\mu=1$
GeV are also shown. It is seen that, at variance with the results obtained
in other works~\cite{Praszalowicz:2001wy}, within our approach the
absolute values of the expansion coefficients $a_{n}$ decrease rather
slowly with $n$.

\begin{table}[tbp] \centering
\begin{tabular}
[c]{|c|c|c|c|c|c|c|}\hline
LO & $a_{2}\left(  \mu\right)  $ & $a_{4}\left(  \mu\right)  $ & $a_{6}\left(
\mu\right)  $ & $a_{8}\left(  \mu\right)  $ & $a_{10}\left(  \mu\right)  $ &
$a_{12}\left(  \mu\right)  $\\\hline
\ $\mu=1%
\operatorname{GeV}%
$ \  & \ $0.0047$ \  & \ $-0.0407$ \  & \ $0.0006$ \  & \ $-0.0185$ \  &
\ $0.0081$ \  & \ $-0.0144$ \ \\
$\mu=2%
\operatorname{GeV}%
$ & $0.0037$ & $-0.0281$ & $0.0004$ & $-0.0112$ & $0.0047$ & $-0.0080$\\
$\mu=3%
\operatorname{GeV}%
$ & $0.0033$ & $-0.0238$ & $0.0003$ & $-0.0089$ & $0.0036$ & $-0.0061$\\\hline
\end{tabular}
\caption{Coefficients of the Gegenbauer expansion calculated at LO}\label{Table_a_1}%
\end{table}%
%

\begin{table}[tbp] \centering
\begin{tabular}
[c]{|c|c|c|c|c|c|c|}\hline
NLO & $a_{2}\left(  \mu\right)  $ & $a_{4}\left(  \mu\right)  $ &
$a_{6}\left(  \mu\right)  $ & $a_{8}\left(  \mu\right)  $ & $a_{10}\left(
\mu\right)  $ & $a_{12}\left(  \mu\right)  $\\\hline
\ $\mu=1%
\operatorname{GeV}%
$ \  & \ $0.0113$ \  & \ $-0.0482$ \  & \ $-0.0019$ \  & \ $-0.0242$ \  &
\ $0.0081$ \  & \ $-0.0189$ \ \\
$\mu=2%
\operatorname{GeV}%
$ & $0.0048$ & $-0.0289$ & $-0.0002$ & $-0.0117$ & $0.0045$ & $-0.0084$\\
$\mu=3%
\operatorname{GeV}%
$ & $0.0033$ & $-0.0238$ & $0.0003$ & $-0.0089$ & $0.0036$ & $-0.0061$\\\hline
\end{tabular}
\caption{Coefficients of the Gegenbauer expansion calculated at NLO.}
\label{Table_a_2}%
\end{table}%
%

\begin{table}[tbp] \centering
\begin{tabular}
[c]{|l|c|c|c|c|c|c|}\hline
& $a_{2}\left(  \mu\right)  $ & $a_{4}\left(  \mu\right)  $ & $a_{6}\left(
\mu\right)  $ & $a_{8}\left(  \mu\right)  $ & $a_{10}\left(  \mu\right)  $ &
$a_{12}\left(  \mu\right)  $\\\hline
Ref.~\cite{Agaev:2012tm} & $0.10$ & $0.10$ & $0.10$ & $0.034$ & 0 & 0\\
Ref.~\cite{Praszalowicz:2001wy} ($M=350$, $n=1$)\ & $0.114$ & $0.015$ & $0.001$ &
$0.0001$ & $-$ & $-$ \\
Ref.~\cite{Praszalowicz:2001wy} ($M=350$, $n=2$)\ & $0.066$ & $-0.027$ & $-0.017$ &
$-0.006$ & $-$ & $-$ \\
Ref.~\cite{Bakulev:2001pa} & $0.20$ & $-0.14$ & $0$ & $0$ & $-$ & $-$ \\
Ref.~\cite{Kroll:2010bf} ($\mu=2$~GeV) & $0.22$ & $0.01$ & $-$ & $-$ & $-$
& $-$ \\\hline
\end{tabular}
\caption{Gegenbauer coefficients for the $\pi$DA given by various authors.
The scale is $\mu=1$~GeV, except in the last row. An exhaustive list of
results is given in Refs.~\cite{Mikhailov:2009kf} and
\cite{Agaev:2010aq}.}
\label{Table_a_3}%
\end{table}%

\section{Light-cone wave function and $k_{T}$ dependence}

\label{Sec_kT}

The concept of $\pi$DA is often associated to that of light-cone wave function
(lcwf). If the pion wave function is expanded in terms of Fock states, the
first (valence) component, dominant at large $Q^{2}$, is the lcwf $\Phi_{\pi
}^{A}\left(  x,k_{T}\right) $, defined by
\begin{eqnarray}
i\sqrt{2}f_{\pi}~\Phi_{\pi}^{A}\left(  x,k_{T}\right)  & = &  \int
\frac{dz^{-}\,d^{2}z_{T}}{2\pi}~e^{iP^{+}z^{-}\left(  x-\frac{1}{2}\right)
-i\vec{k}_{T}\cdot\vec{z}_{T}}\,\times\nonumber\\
& & \left.  \left\langle 0\right\vert \bar{u}\left(  -\frac{z}{2}\right)
\gamma^{+}\gamma_{5}\,d\left(  \frac{z}{2}\right)  \left\vert \pi^{-}\left(
P\right)  \right\rangle \right\vert _{z^{+}=0}\ .\label{wf.1}%
\end{eqnarray}
The label $A$, denoting \textquotedblleft axial\textquotedblright%
\ lcwf, has been used, e.g., in Ref.~\cite{Zhitnitsky:1993vb}. When
dealing with hard-exclusive processes, the lcwf Eq.~(\ref{wf.1}),
integrated with respect to $k_{T}$, can be identified with the $\pi$DA
\cite{sterman}. Therefore, in order to carry out a phenomenological
analysis, some authors do not distinguish between the ($k_{T}$-integrated)
lcwf and the $\pi$DA. In this section we compare the predictions in those
works with ours, paying special attention to the results related to the
quark transverse momentum $k_{T}$.
To this aim, some caveats are in order.

The most direct comparison that could be performed is that between the
results obtained in other works for $\Phi_{\pi}^{A}(x,k_{T})$,
Eq.~(\ref{wf.1}), and those obtained here for the $\pi$DA,
Eq.~(\ref{DA.01}). If we write
\begin{equation}
\phi_{\pi}(x)  =\int\frac{d^{2}k_{T}}{\left(  2\pi\right)  ^{2}%
}\; \Phi_\pi(x,k_{T})  ~, \label{wf.21a}%
\end{equation}
it is natural to identify [c.f.~Eq.~(\ref{DA.01})]
\begin{equation}
\Phi_{\pi}^{A}(x,k_{T})  \equiv\Phi_{\pi}(  x,k_{T})
=\frac{2\sqrt{2}N_{c}\,g_{\pi q\bar q}}{f_{\pi}}\int\frac{dw}{\left(  2\pi\right)
^{2}}\,F\left(  w,x,k_{T}\right)  ~. \label{wf.19}%
\end{equation}
On the other hand, in some works the lcwf has been identified with a
different quantity, which in our context would correspond to
$\Phi_{\pi}^{(0)}(x,k_{T})$, obtained from the relation
\begin{equation}
\phi_{\pi}^{(0)}\left(  x\right)  =\int\frac{d^{2}k_{T}}{\left(  2\pi\right)
^{2}} \; \Phi_{\pi}^{(0)}\left(  x,k_{T}\right)  ~, \label{wf.2}%
\end{equation}
with $\phi_{\pi}^{(0)}(x)$ given by Eq.~(\ref{DA.17}). From
Eqs.~(\ref{DA.17}) and (\ref{wf.2}), $\Phi_{\pi}^{(0)}(x,k_{T})$ can be
cast in the form
\begin{equation}
\Phi_{\pi}^{\left(  0\right)  }\left(  x,k_{T}\right)  =N\int dw\,\frac
{g_{k}\,z_{k_{+}}\,z_{k_{-}}\,\left[  \left(  1-x\right)  \,m_{k_{+}%
}+x\,m_{k_{-}}\right]  }{\left(  k_{+}^{2}+m_{k_{+}}^{2}\right)  \,\left(
k_{-}^{2}+m_{k_{-}}^{2}\right)  }\ , \label{wf.01}%
\end{equation}
where
$N$ is a normalization factor.

We recall that $\phi_{\pi}^{(0)}(x)$ is the $\pi$DA evaluated in nlNJL
models using the operator $\gamma^{+}\gamma_{5}$ without dressing, while
the full $\pi$DA obtained in the present nlNJL approach includes also
other operators carrying different tensor structures, namely
$\bar{u}\left(  p_{1}+p_{2}\right)  ^{+}\left(  \not p _{1}+\not p
_{2}\right)  \gamma_{5}d$ and $\bar{u}\left(  p_{1}-p_{2}\right)
^{+}\gamma_{5}d,$ where $p_{1,2}$ are the quark momenta. We emphasize
therefore that in the present scheme the ($k_{T}$-integrated) pion lcwf
and the $\pi$DA are different quantities. In particular, as it is
discussed in the previous section, the latter satisfies exactly the
normalization sum rule Eq.~(\ref{PionDA.05}).

Thus, in the following we will compare $\Phi_{\pi}^{A}(x,k_{T})$,
evaluated within other approaches, with our results for both the
quantities $\Phi_{\pi}(x,k_{T})$ and $\Phi_{\pi}^{(0)}(x,k_{T})$. It is
worth stressing that some predictions concerning the $k_{T}$ dependence
could be ultimately related to observables.

Let us consider the quantities
\begin{equation}
\left\langle k_{T}^{2}\right\rangle _{(0)}=\frac{\int dx\,d^{2}k_{T}%
\,k_{T}^{2}\,\left\vert \Phi_{\pi}^{\left(  0\right)  }\left(  x,k_{T}\right)
\right\vert ^{2}}{\int dx\,d^{2}k_{T}\,\left\vert \Phi_{\pi}^{\left(
0\right)  }\left(  x,k_{T}\right)  \right\vert ^{2}}~ \label{wf.02}%
\end{equation}
and
\begin{equation}
\left\langle k_{T}^{2}\right\rangle = \frac{\int dx\,d^{2}k_{T}\,k_{T}^{2}
\,\left\vert \Phi_{\pi}\left(  x,k_{T}\right)  \right\vert ^{2}} {\int
dx\,d^{2}k_{T}\,\left\vert \Phi_{\pi} \left(  x,k_{T} \right)  \right\vert
^{2}}~. \label{wf.03}%
\end{equation}
Since $\Phi_{\pi}^{(0)}\left(  x,k_{T}\right)  $ and $\Phi_{\pi}\left(
x,k_{T}\right)  $ are probability amplitudes, either $\left\langle k_{T}%
^{2}\right\rangle $ or $\left\langle k_{T}^{2}\right\rangle _{(0)}$ can be
interpreted as the average transverse momentum of the valence quark. For
the region of high $Q^{2}$ (i.e., where the lcwf is thought to be the dominant
contribution to the pion wave function~\cite{sterman}) this quantity could be
accessed in future measurements, performed along the lines proposed in
Ref.~\cite{aitala}.

In our framework we get $\left\langle k_{T}^{2}\right\rangle _{(0)}^{1/2}=270$
MeV and $\left\langle k_{T}^{2}\right\rangle ^{1/2}=260$ MeV. It is
interesting to compare these values with the result obtained in
Ref.~\cite{Kroll:2010bf}, namely $\left\langle k_{T}^{2}\right\rangle
^{1/2}\simeq$ 710 MeV, where the average is evaluated considering an axial
pion lcwf at a scale $\mu=1$ GeV. The corresponding $k_{T}$ dependence is
given by
\begin{equation}
\Phi_{\pi}^{A}\left(  x,k_{T}\right)  =\phi_{\pi}\left(  x\right)
\,\frac{4\,\pi\,\sigma_{\pi}^{2}}{x\,\left(  1-x\right)  }\exp\left(
-\frac{k_{T}^{2}\,\sigma_{\pi}^{2}}{x\,\left(  1-x\right)  }\right)  ~,
\label{kroll}%
\end{equation}
where $\phi_{\pi}\left(  x\right)  $ is the $\pi$DA. Although at first
sight the results seem to disagree, if we make use of Eqs.~(8) and (9) of
Ref.~\cite{Kroll:2010bf} in order to determine the value of $\left\langle
k_{T}^{2}\right\rangle ^{1/2}$ at the asymptotic limit, and take for the
``traverse size parameter'' the value $\sigma_{\pi}\sim 1$~GeV$^{-1}$
(upper limit of the range considered in Ref.~\cite{Kroll:2010bf}), we get
$\left\langle k_{T}^{2}\right\rangle ^{1/2}\sim300$~MeV. Therefore, our
result is found to be somewhat lower but not incompatible with that
obtained in Ref.~\cite{Kroll:2010bf}.

Now let us also consider a pseudoscalar pion lcwf, $\Phi_{\pi}^{P}\left(  x,
k_{T} \right)  $. The latter has been introduced in
Ref.~\cite{Zhitnitsky:1993vb},
with the aim of obtaining constraints on the lcwf in a light-cone sum rule
framework. In order to analyze this function in the context of the nlNJL
model, let us start by defining the $k_{T}$-integrated pseudoscalar function
$\phi_{\pi}^{P}(x)$, which is higher twist with respect to the axial one:
\begin{equation}
-\frac{2\sqrt{2}\left\langle \bar{q}q\right\rangle }{P^{+}\,f_{\pi}}~\phi
_{\pi}^{P}\left(  x\right)  =\int\frac{dz^{-}}{2\pi}~e^{iP^{+}z^{-}\left(
x-\frac{1}{2}\right)  }~\left.  \left\langle 0\right\vert \bar{\psi}\left(
-\frac{z}{2}\right)  i\gamma_{5}\tau^{-}\psi\left(  \frac{z}{2}\right)
\left\vert \pi^{-}\left(  P\right)  \right\rangle \right\vert _{z^{+}%
=0,\vec{z}_{\bot}=0}~. \label{wf.10}%
\end{equation}
It is seen that $\phi_{\pi}^{P}(x)$ fulfills an approximate sum rule, which
becomes exact in the chiral limit:%
\begin{equation}
\int_{0}^{1}dx\,\phi_{\pi}^{P}\left(  x\right)  =1+\mathcal{O}\left(  m_{\pi
}^{2}\right)  ~. \label{wf.13}%
\end{equation}
For this function, within the present approach one gets the result%
\begin{eqnarray}
\phi_{\pi}^{P}\left(  x\right) & = & \int\frac{d^{2}k_{T}}{\left(  2\pi\right)
^{2}}\;\Phi^{P}_{\pi}\left(  x,k_{T}\right) \nonumber\\
& = & -\frac{f_{\pi}\,\sqrt{2}N_{c}%
}{\left\langle \bar{q}q\right\rangle }\;g_{\pi q\bar q}\int\frac{dw\,d^{2}k_{T}%
}{\left(  2\pi\right)  ^{4}}\;\frac{g_{k}\,z_{k_{+}}\,z_{k_{-}}\left(
k^{2}+\frac{m_{\pi}^{2}}{4}+m_{k_{+}}m_{k_{-}}\right)  }{\left(  k_{+}%
^{2}+m_{k_{+}}^{2}\right)  \,\left(  k_{-}^{2}+m_{k_{-}}^{2}\right)  }~,
\label{wf.11}%
\end{eqnarray}
hence the pseudoscalar lcwf will be given by
\begin{equation}
\Phi^{P}_{\pi}\left(  x,k_{T}\right)  =-\frac{f_{\pi}\,\sqrt{2}N_{c}%
}{\left\langle \bar{q}q\right\rangle }\;g_{\pi q\bar
q}\int\frac{dw}{\left(  2\pi\right)
^{2}}\;\frac{g_{k}\,z_{k_{+}}\,z_{k_{-}}\,\left(
k^{2}+\frac{m_{\pi}^{2}}{4}+m_{k_{+}}m_{k_{-}}\right)  }{\left(  k_{+}%
^{2}+m_{k_{+}}^{2}\right)  \,\left(  k_{-}^{2}+m_{k_{-}}^{2}\right)  }~.
\label{wf.20}%
\end{equation}
In addition, with the aim of finding light-cone sum rules, in
Refs.~\cite{Zhitnitsky:1993vb,Praszalowicz:2001wy} the authors also
consider the following $k_{T}$ moments of the lcwf $\Phi^{A,P}_\pi
(x,k_{T})$:
\begin{equation}
\left\langle k_{T}^{m}\right\rangle _{A,P}=\int dx\,\frac{d^{2}k_{T}}{\left(
2\pi\right)  ^{2}}\,k_{T}^{m}\,\Phi_{\pi}^{A,P}\left(  x,k_{T}\right)
\ ,\quad\mbox{with $m = 2,4$}\ . \label{wf.21}%
\end{equation}
It is important to remark that $\Phi_{\pi}^{A,P}\left(  x,k_{T}\right)  $
are not momentum density distributions. In fact, there is no guarantee
that these functions are positive defined. Therefore Eq.~(\ref{wf.21}) is
not related to observable quantities and it may be useful only for
theoretical considerations. In the present approach, from the analytical
expressions in Eqs.~(\ref{wf.19}) and (\ref{wf.20}) it is seen that for
large $k_{T}$ the functions $\Phi_{\pi}^{A,P}\left( x,k_{T}\right)  $
behave as
\begin{align*}
&  \Phi_{\pi}^{A}\left(  x,k_{T}\right)  \underset{k_{T}\rightarrow\infty
}{\longrightarrow}k_{T}^{-5}\ ,\\
&  \Phi_{\pi}^{P}\left(  x,k_{T}\right)  \underset{k_{T}\rightarrow\infty
}{\longrightarrow}k_{T}^{-3}\ .
\end{align*}
In view of these asymptotic behaviors, only the estimate $\left\langle
k_{T}^{2}\right\rangle _{A}^{1/2}=445$~MeV can be obtained, while
$\left\langle k_{T}^{4}\right\rangle _{A}$ and $\langle k_{T}^{2,4}\rangle
_{P}$ are not well defined. One has to say that, in our approach, only the
nonperturbative $k_{T}$ dependence arises naturally from the model
calculation. As it is well known, an additional perturbative dependence is
found if one takes into account that configurations with two quarks
carrying high $k_{T}$ are suppressed due to gluon
radiation~\cite{Lepage:1979zb}. This is the origin, for example, of the
factor $k_{T}^{2} \sigma_{\pi}^{2}$ in the exponent of Eq.~(\ref{kroll}).
We have considered the possibility of obtaining a prediction for
$\left\langle k_{T}^{4}\right\rangle _{A}$ and $\langle
k_{T}^{2,4}\rangle_{P}$ by including a high $k_{T}$ suppression factor
such as e.g.\ these exponential functions in our lcwfs. However, we have
found that our results are quite sensitive to the cutoff prescription,
hence we are not able to provide a robust prediction for these quantities.

\begin{figure}
[ptb]
\begin{center}
\includegraphics[
height=2.7256in,
width=3.8696in
]%
{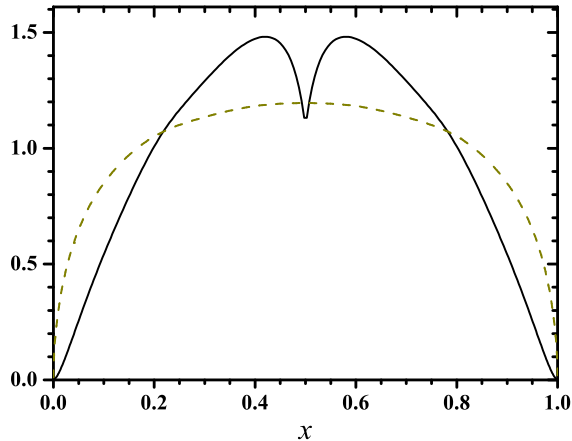}%
\caption{$\pi$DA (solid line) and light-cone wave function
$\phi_{\pi}^{P}(x)$ (dashed line), Eq.~(\ref{wf.10}), in the nlNJL
model.}
\label{Fig_DA_model_B}%
\end{center}
\end{figure}

\begin{figure}
[ptb]
\begin{center}
\includegraphics[
width=4.7in
]%
{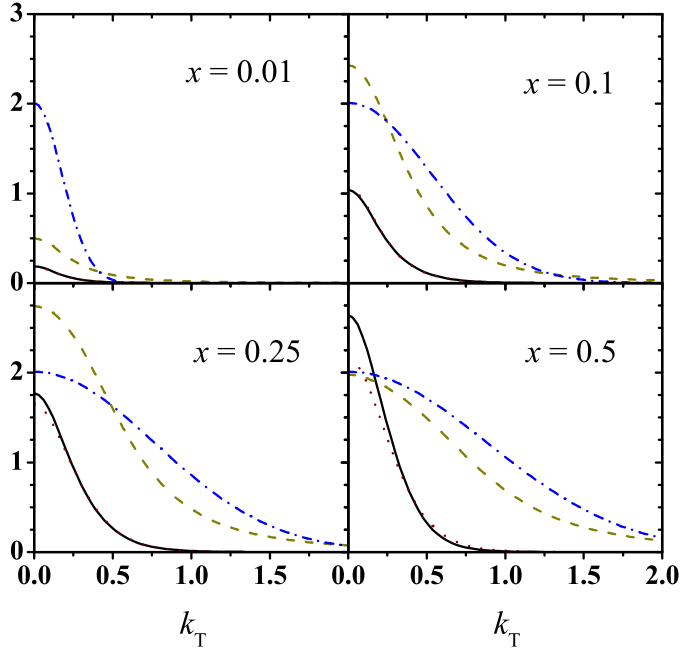}%
\caption{Solid and dotted lines correspond to the pion axial lcwf $\Phi_{\pi
}^{A}(x,k_{T})$, Eq.~(\ref{wf.19}), and the function $\Phi_{\pi}^{(0)}\left(
x,k_{T}\right)  $, Eq.~(\ref{wf.01}), respectively, both multiplied by a
factor $x(1-x)/[10 \, \phi_{\pi}(x)]$. The dashed line stands for the pion
pseudoscalar lcwf $\Phi_{\pi}^{P}\left(  x,k_{T}\right)  $, Eq.~(\ref{wf.20}),
while the dashed-dotted line corresponds to the pion axial lcwf considered in
Ref.~\cite{Kroll:2010bf}, Eq.~(\ref{kroll}), both multiplied by $x(1-x)/\phi
_{\pi}(x)$. We remark that the results for the first two quantities are
rescaled by a factor of 10 with respect to the last two.}
\label{Fig_lcwf_kT}%
\end{center}
\end{figure}

It is worth noticing that the wave functions $\Phi_{\pi}^{P}\left(
x,k_{T}\right)  $ and $\Phi_{\pi}^{A}\left(  x,k_{T}\right)  $ are quite
different from each other. From Fig.~\ref{Fig_DA_model_B}, where we have
plotted $\phi_{\pi}^{P}\left(  x\right)$ together with $\phi_{\pi}%
^{A}\left(  x\right)  \equiv\phi_{\pi}\left(  x\right)  $, we observe that
$\phi_{\pi}^{P}\left(  x\right)  $ has less structure than
$\phi_{\pi}^{A}(x)$. In fact, $\phi_{\pi}^{P}(x)$ appears to be close to a
flat distribution, which corresponds to the asymptotic limit
$\phi_{\pi}^{P}(x) = 1$. Moreover, the $k_{T}$ dependence is also very
different. This can be seen in Fig.~\ref{Fig_lcwf_kT}, where we show our
results for the functions $\Phi_{\pi}^{A}(x,k_{T})$,
$\Phi_{\pi}^{P}(x,k_{T})$ and $\Phi_{\pi}^{(0)}(x,k_{T})$ as functions of
$k_{T}$ for some definite values of $x$. We also include the results for
the axial pion lcwf proposed in
Ref.~\cite{Kroll:2010bf}, Eq.~(\ref{kroll}), with $\sigma_{\pi}=0.4$%
~GeV$^{-1}$. In the figure, the results are presented in such a way that
the curves corresponding to the lcwf in Eq.~(\ref{kroll}) have the same
value at $k_{T}=0$ for all values of $x$. It is clear that neither the
shape nor the size of our functions $\Phi_{\pi}^{A}\left(  x,k_{T}\right)
$ and $\Phi_{\pi }^{(0)}\left(  x,k_{T}\right)  $ support the $k_{T}$
dependence proposed in Ref.~\cite{Kroll:2010bf}. Instead, the latter shows
a somewhat qualitative agreement with our results for
$\Phi_{\pi}^{P}\left(  x,k_{T}\right)  $, at
least, for values of $x$ above say 0.1.%

\section{The pion-photon transition form factor}

\label{Sec_TFF}

In this section we present the results for the $\pi-\gamma$ TFF obtained
within our approach, i.e.~via the $\pi$DA described in Sec.~III. Here we
have modified the expression in Eq.~(\ref{PionTFF.01}) by adding
sub-leading terms in the expansion in powers of $1/Q^{2}$. This procedure
has been already used in Ref.~\cite{Noguera:2010fe} in order to simulate
e.g.\ contributions coming from higher twist operators. We propose to
include two additional terms in the expansion, writing
\begin{equation}
Q^{2}\,F(  Q^{2}) \ = \ \frac{\sqrt{2}f_{\pi}}{3}\left[  \int_{0}%
^{1}dx\,T_{H}\left(  x,Q^{2},\mu\right)  \,\phi_{\pi}(x,\mu)
+\frac{C}{Q^{2}}+\frac{D}{Q^{4}}\right]  \ ,
\label{TFF.02}%
\end{equation}
where $C$ and $D$ are constants to be determined by fitting our expression
to the experimental data. For the scale $\mu$ we will take $\mu^2 = Q^2$.

\begin{figure}
[ptb]
\begin{center}
\includegraphics[
width=5in 
]%
{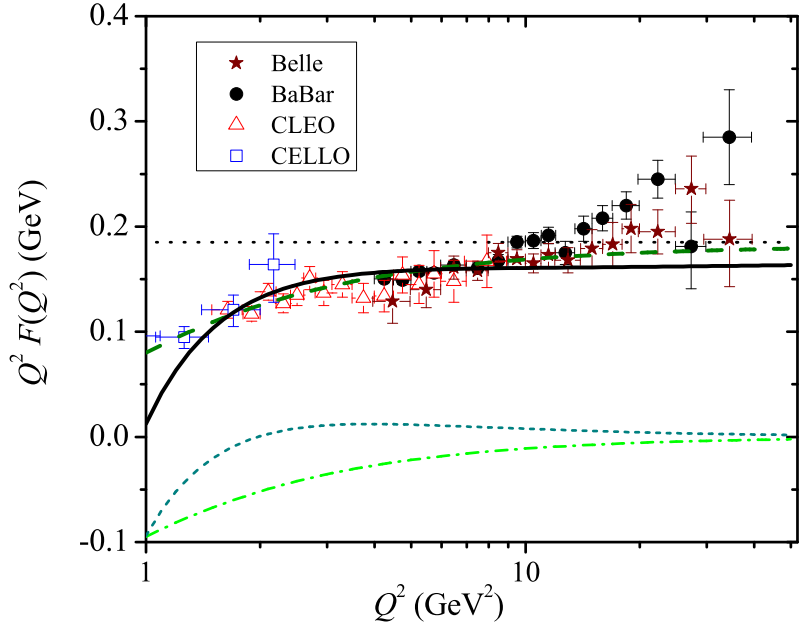}%
\caption{Values of $Q^{2}\, F(Q^{2})$, Eq.~(\ref{TFF.02}), at LO
(long-dashed line) and NLO (solid line), in comparison with experimental
data. Dashed-dotted and short-dashed curves show the contributions given
by the term $C/Q^{2}+D/Q^{4}$ at LO and NLO, respectively, while the
horizontal dotted line indicates the asymptotic QCD limit.}
\label{Fig_Pi_Gam_Gam_setC}%
\end{center}
\end{figure}

In Fig.~\ref{Fig_Pi_Gam_Gam_setC} we show our results for $Q^{2}\,F(Q^{2})$.
Long-dashed and solid curves correspond to LO and NLO evolutions of the
$\pi$DA, respectively, whereas the dashed-dotted (short-dashed) line stands
for the contribution of the term $C/Q^{2}+D/Q^{4}$ at LO (NLO). The values
obtained for the parameters $C$ and $D$ from a fit to all available data
(i.e.\ including the data from CELLO, CLEO, BABAR and BELLE experiments), up
to LO and NLO accuracy, are listed in the first row of Table~\ref{tabella}.
Only the three data with $Q^2 > 1$~GeV$^2$ of the CELLO Collaboration have
been retained in our fit.

\begin{table}[tbp] \centering
\begin{tabular}
[c]{|c|c|c|c|c|c|c|c|}\hline $\pi$DA & Data set & Accuracy & $C$
[GeV$^{2}$] & $D$ [GeV$^{4}$] & M [GeV] & n$^{\circ}$ points & $\chi^{2}
/$n$^\circ$ points\\\hline
nlNJL & Cello+Cleo+Belle+Babar & LO & -1.82 & 0.29 & - & 50 & 1.9 \\
&  & NLO & 1.56 & -3.09 & - & 50 & 3.5\\\hline
nlNJL & Cello+Cleo+Babar & LO & -1.80 & 0.26 & - & 35 & 2.4\\
&  & NLO & 1.49 & -2.95 & - & 35 & 4.3\\\hline
nlNJL & Cello+Cleo+Belle & LO & -2.01 & 0.65 & - & 33 & 0.61\\
&  & NLO & 0.90 & -1.91 & - & 33 & 1.09\\\hline
flat & Cello+Cleo+Belle+Babar & LO & 1.82 & -1.50 & 0.76 & 50 & 0.91\\
&  & NLO & 1.47 & -1.08 & 0.57 & 50 & 0.96\\\hline
\end{tabular}
\caption{Values of the parameters $C$, $D$ and $M$ [see
Eqs.~(\ref{TFF.02}) and (\ref{TFF.07})], obtained from the fits to
different experimental data sets for the $\pi$TFF. The first three rows
correspond to the $\pi$DA calculated within the nlNJL model, while entries
in the last row are obtained from a flat $\pi$DA.}
\label{tabella}
\end{table}

Three main conclusions can be outlined from these results: $i)$ the
overall agreement between the fitted curve and the data is not
satisfactory; $ii)$ the values of the parameters $C$ and $D$ are not
stable when going from LO to NLO; $iii)$ what is more relevant, the
accuracy of the fit is rather worse at NLO than at LO.

As it has been done in Ref.~\cite{Bakulev:2011rp}, we have also considered
separately the inclusion of the data from BABAR and those from BELLE. From
Table~\ref{tabella} (second row) it is seen that if one excludes the BELLE
results from the full data set, the picture does not change appreciably.
On the other hand, if one excludes the BABAR data the situation is somehow
different (see third row in Table~\ref{tabella}): while the agreement with
the data gets improved, problems $ii)$ and, in particular, $iii)$, still
remain. One can therefore conclude that BELLE data can be easily
adjustable in our scheme, especially at LO, and that the corrections
arising from NLO contributions to the evolution equations go in the wrong
direction, in all cases under study.

In order to test this last statement, we have checked what happens if,
instead of the $\pi$DA from our nlNJL model, we take as input a flat
distribution $\phi_{\pi}\left( x\right)=1$. In order to avoid
singularities, in this case we modify the kernel $T_{H}\left(
x,Q^{2},\mu\right)$ introducing a new parameter
$M$~\cite{Radyushkin:2009zg}:
\begin{equation}
T_{H}^{\text{NLO}}\left(  x,Q^{2},\mu\right)  =\frac{1}{x+\frac{M^{2}}{Q^{2}}%
}\left\{  1+C_{F}\frac{\alpha_{\mathrm{s}}\left(  \mu\right)  }{4\,\pi}\left[
\ln^{2}x-\frac{x\,\ln x}{1-x}-9+\left(  3+2\ln x\right)  \ln\frac{Q^{2}}%
{\mu^{2}}\right]  \right\} \ . \label{TFF.07}%
\end{equation}
We take here the scale $\mu_0=1$~GeV~\cite{Noguera:2010fe}, at which one
assumes that the quark model provides a good description of low energy
physics.

\begin{figure}
[ptb]
\begin{center}
\includegraphics[
width=5in 
]%
{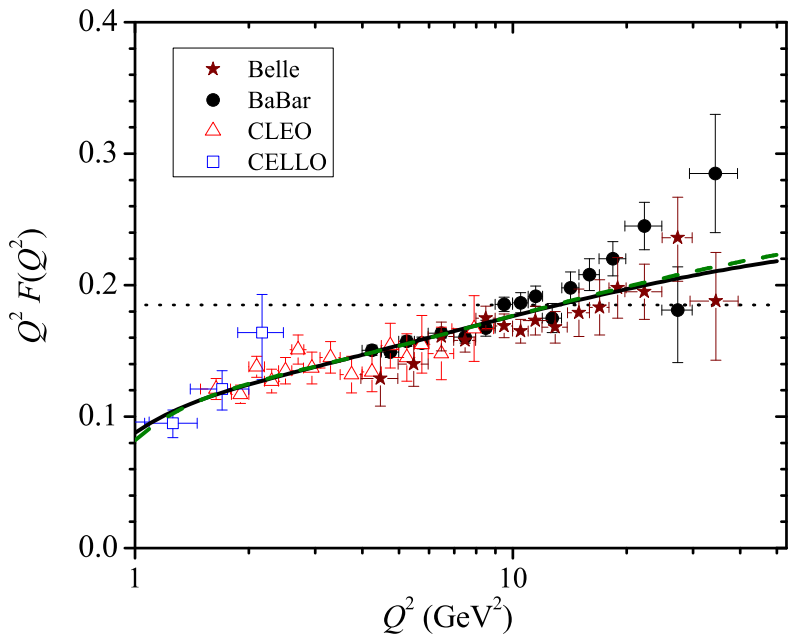}%
\caption{Values of $Q^{2}\,F(Q^{2})$ obtained from a flat pion
distribution amplitude, at LO (dashed line) and NLO (dashed line), in
comparison with experimental data. The horizontal dotted line indicates
the asymptotic QCD limit.}
\label{Fig_Pi_Gam_Gam_NJL}%
\end{center}
\end{figure}

In Fig.~\ref{Fig_Pi_Gam_Gam_NJL} (see also fourth row of
Table~\ref{tabella}) we show the results for $Q^{2}\,F(Q^{2})$ obtained
after inserting the function in Eq.~(\ref{TFF.07}) into
Eq.~(\ref{TFF.02}), for a flat distribution $\phi_{\pi}\left( x\right)=1$
[at the LO, only the first term into the brackets in Eq.~(\ref{TFF.07})
has to be considered]. It is seen that the agreement with the full set of
experimental data becomes improved with respect to the previous analyses
(one should notice anyway that a further parameter, $M$, has been
included), and that the parameters of the fit are more stable when passing
from LO to NLO. We stress, however, that the important conclusion $iii)$
stated in the previous cases still holds: the inclusion of NLO corrections
does not help to describe the experimental data. This becomes more evident
for virtualities $Q^{2}$ above $10$~GeV$^{2}$.

Due to the regularization of $T_H$ in the limit $x\to 0$, it is difficult
to compare our results for the flat distribution with those obtained
within approaches based on the parametrization of the Gegenbauer
expansion. For instance, one could define an effective $\pi$DA
$\phi_{\pi}^{\rm eff}(x,Q)$ by the relation
\[
\left(  \frac{1}{x}+\frac{1}{1-x}\right)  \phi_{\pi}^{\rm eff}\left(
x,Q\right)
=\left(  \frac{1}{x+\frac{M^{2}}{Q^{2}}}+\frac{1}{1-x+\frac{M^{2}}{Q^{2}}%
}\right)  \phi_{\pi}\left(  x,Q\right)\ ;
\]
now the problem is that it would not be guaranteed that $a_{0}(Q) = 1$ in
the Gegenbauer expansion, Eq.~(\ref{DA.31}). Actually, if we assume a flat
distribution $\phi_\pi(x,\mu_0) = 1$, we get
\[
\int_{0}^{1}dx\,\phi_{\pi}^{\rm eff}\left(  x,\mu_{0}\right)  =1+2\frac{M^{2}}%
{\mu_{0}^{2}}-2\frac{M^{2}}{\mu_{0}^{2}}\left(
1+\frac{M^{2}}{\mu_{0}^{2}}\right) \log\left(
1+\frac{\mu_{0}^{2}}{M^{2}}\right)\ ,
\]
which is equal to 0.33 for $M=0.76$~GeV and to 0.44 for $M=0.58$~GeV.
Therefore, we can not compare the coefficients in the Gegenbauer expansion
with those obtained in Tables~\ref{Table_a_1}, \ref{Table_a_2} and
\ref{Table_a_3}.

Finally, it is interesting to notice that the conclusion concerning the
NLO corrections is also valid for the asymptotic behavior of the $\pi$DA.
Indeed, taking $\phi_{\pi}\left(  x\right) =6\,x\,(1-x)$ we find, at the
LO, $Q^2\,F\left( Q^{2}\right) =\sqrt{2}f_{\pi}=0.185$~GeV, and at the
NLO, $Q^{2}\,F\left( Q^{2}\right) =\sqrt{2}f_{\pi}\left[
1-0.53\,\alpha_{\mathrm{s}}\left( Q^{2}\right)
\right]  =0.161 (0.164)$~GeV 
for $Q^{2}=10(20)$~GeV$^{2}$. Therefore, the NLO correction reduces the
$\pi$TFF by about a 13\%, in a direction which is opposite to that
of the data.

\section{Conclusions}

\label{Sec_Conclusion}

In this work, the $\pi$DA and the associated $\pi$TFF have been evaluated
within the framework of a nonlocal Nambu--Jona-Lasinio model that has been
shown to succesfully describe several pion
observables~\cite{Noguera:2008,Dumm:2010hh}. In this approach, the
couplings between nonlocal quark currents ensure the preservation of
chiral, Poincar\'{e} and local electromagnetic gauge invariances. The
three main ingredients of the calculation are the description of the pion
as a bound state of a Bethe-Salpeter equation, the usage of a prescription
for the introduction of the electroweak interaction vertices and,
eventually, the quark propagator, which shows the momentum dependence
obtained in lattice QCD. The calculated $\pi$DA has to be therefore
associated to the momentum scale of the lattice data, namely
3~GeV~\cite{Parappilly:2005ei}. In general, the comparison of any
observable related to the $\pi$DA (as e.g.\ the $\pi$TFF) with
experimental data will require a perturbative evolution of the results
obtained at this reference scale. Here this evolution has been carried out
up to NLO accuracy.

Since the theoretical framework respects all basic symmetries, our $\pi$DA
is shown to fulfill three fundamental properties: it has the correct
symmetry in the quark momentum fraction, it is defined in the proper
support and, above all, it turns out to be naturally normalized, a feature
which is imposed in other
schemes~\cite{Praszalowicz:2001wy,Dorokhov:2013xpa,
Bakulev:2001pa,Mikhailov:2009kf,Mikhailov:2010ud,Bakulev:2011rp,
Agaev:2010aq,Agaev:2012tm,Huang:2013yya,Kroll:2010bf}. It is seen that our
$\pi$DA, already at the scale of 3~GeV, is not far from the asymptotic
distribution $\phi_\pi(x)=6x(1-x)$. In fact, we find that the genuine
nonlocal contributions push the result towards this asymptotic behavior.
Moreover, the pseudoscalar pion distribution amplitude is also found to
approach its corresponding asymptotic limit $\phi_\pi^P(x)= 1$. Another
outcome of our results is that when the $\pi$DA is expanded in Gegenbauer
polynomials, in contrast with other
calculations~\cite{Praszalowicz:2001wy,Agaev:2010aq,Agaev:2012tm} we find
that the absolute values of the corresponding coefficients $a_n$ decrease
rather slowly with $n$.

The last part of the paper is devoted to phenomenological considerations.
Our results for the functions $\Phi_\pi(x,k_T)$, where $k_T$ is the quark
transverse momentum, are compared to those obtained within a light-cone
wave function approach. It is found that the $k_{T}$ dependence obtained
in our framework turns out to be rather different from that calculated in
other works~\cite{Kroll:2010bf}. This feature could in principle be
checked in future experiments. Concerning the evaluation of the $\pi$TFF,
we have found that NLO corrections in general lead to a suppression of
$Q^2F(Q^2)$, which represents a problem towards the explanation of the
already challenging experimental scenario. In particular, in our nlNJL
approach (which is based on the evaluation of standard diagrams, and
considers just general assumptions such as chiral symmetry and lattice
results), it is very problematic to obtain a $\pi$TFF that crosses the
aymptotic limit as suggested by the pattern of the BABAR data.

\section*{Acknowledgements}

We thank A.\ Pimikov for a critical reading of the manuscript. This work has
been partially funded by the Spanish MCyT (and EU FEDER) under contract
FPA2010-21750-C02-01 and AIC10-D-000588, by Consolider Ingenio 2010 CPAN
(CSD2007-00042), by Generalitat Valenciana: Prometeo/2009/129, by the
European Integrated Infrastructure Initiative HadronPhysics3 (Grant number
283286), by CONICET (Argentina) under grants \# PIP 00682 and PIP 02495, and
by ANPCyT (Argentina) under grant \# PICT-2011-0113.

\bigskip

\appendix{}

\section{Derivation of the $\pi$DA in the nonlocal NJL model}

\label{App_NL_NJL} In this Appendix we provide some details on the
obtention of the $\pi$DA in Eq.~(\ref{DA.01}). We start with the Euclidean
action in Eq.~(\ref{action}), and include a coupling with an external
axial gauge field $a_{\mu}$, as described in Sect.~IIB. In order to deal
with meson degrees of freedom, it is convenient to bosonize the fermionic
theory by introducing scalar and pseudoscalar fields $\sigma_{1,2}(y)$ and
$\vec{\pi}(y)$ and integrating out the fermion fields. This bosonized
action can be written as~\cite{Noguera:2008,Dumm:2010hh}
\begin{equation}
S^{\mathrm{bos}}=-\ln\det\mathcal{D}+\frac{1}{2G_{S}}\int
d^{4}y\Big[\sigma
_{1}(y)\sigma_{1}(y)+\sigma_{2}(y)\sigma_{2}(y)+\vec{\pi}(y)\cdot
\vec{\pi}(y)\Big]
\ ,
\end{equation}
where
\begin{eqnarray}
\!\!\!
\mathcal{D}\left(  y+\frac{z}{2},y-\frac{z}{2}\right)   &  = &\gamma
_{0}\;W\left(  y+\frac{z}{2},y\right)  \gamma_{0}\Bigg\{\delta^{(4)}(z)\left[
-i\rlap/\partial+m_{c}\right]  +\nonumber\\
& & \left[  \mathcal{G}(z)\left[  \sigma_{1}\left(  y\right)
+i\vec{\tau}\cdot\vec{\pi}\left(  y\right)  \right]  +\mathcal{F}%
(z)\ \sigma_{2}\left(  y\right)  \frac{i{\overleftrightarrow{\rlap/\partial}}%
}{2\ \varkappa_{p}}\right]  \Bigg\}W\left(  y,y-\frac{z}{2}\right)  \label{aa}%
\end{eqnarray}
As usual we assume that the fields $\sigma_{1,2}$ have nontrivial
translational invariant mean field values $\bar{\sigma}_{1}$ and
$\varkappa_{p}\,\bar{\sigma}_{2}$, while the mean field values of pseudoscalar
fields $\pi_{i}$ are zero. Thus we write
\begin{equation}
\sigma_{1}(y)=\bar{\sigma}_{1}+\delta\sigma_{1}(y)\ ,\qquad\sigma
_{2}(y)=\varkappa_{p}\,\bar{\sigma}_{2}+\delta\sigma_{2}(y)\ ,\qquad\vec{\pi
}(y)=\delta\vec{\pi}(y)\ .
\end{equation}
Replacing in the bosonized effective action and expanding in powers of meson
fluctuations and the external field $a_{\mu}$ we obtain
\begin{equation}
S^{\mathrm{bos}}=S_{\mathrm{{\small MFA}}}+S_{\mathrm{quad}}+S_{\pi a}%
+\dots\ , \label{sbos}%
\end{equation}
where only the terms relevant for our calculation have been explicitly
written. Here the mean field action per unit volume reads
\begin{equation}
S_{\mathrm{{\small MFA}}}=\frac{1}{2G_{S}}\left(  \bar{\sigma}_{1}%
^{2}+\varkappa_{p}^{2}\ \bar{\sigma}_{2}^{2}\right)  -4N_{c}\int\frac{d^{4}%
p}{(2\pi)^{4}}\ \ln\mathcal{D}_{0}\ ,
\end{equation}
with $\mathcal{D}_{0}=(-\rlap/p+m_{p})/z_{p}\,$, see Eqs.~(\ref{qprop}) and
(\ref{zm}) in Sect.~IIB.

The minimization of $S_{\mathrm{{\small MFA}}}$ with respect to $\bar{\sigma
}_{1,2}$ leads to the corresponding Dyson-Schwinger equations, which together
with Eqs.~(\ref{zm}) and (\ref{parametrization_set2}) allow to determine the
values of $G_{S}$ and $\varkappa_{p}\,$. The quadratic piece of the bosonic
Euclidean action can be written as
\begin{equation}
S_{E}^{\mathrm{quad}}=\frac{1}{2}\int\frac{d^{4}p}{(2\pi)^{4}}\sum
_{M=\sigma,\sigma^{\prime},\pi}G_{M}(p^{2})\ \delta M(p)\ \delta M(-p)\ ,
\label{quad}%
\end{equation}
where the fields $\sigma$ and $\sigma^{\prime}$ are scalar meson mass
eigenstates, defined in such a way that there is no $\sigma-\sigma^{\prime}$
mixing at the level of the quadratic action. The explicit expressions for the
one-loop integrals $G_{M}(p^{2})$, as well as those of the above mentioned
Dyson-Schwinger equations, can be found in Ref.~\cite{Noguera:2008}. Meson
masses can be obtained by solving the associated Bethe-Salpeter equations
$G_{M}(-m_{M}^{2})=0$, while on-shell meson-quark coupling constants
$g_{Mq\bar q}$ are given by
\begin{equation}
{g_{Mq\bar q}}^{-2}=\ \frac{dG_{M}(p^{2})}{dp^{2}}\bigg|_{p^{2}=-m_{M}^{2}%
}\ . \label{gpiqq}%
\end{equation}
Finally, the bilinear piece in $\delta\pi$ and $a_{\mu}$ fields $S_{\pi a}$ in
Eq.~(\ref{sbos}) reads
\begin{equation}
S_{\pi a}=\mathrm{Tr}\left[  \mathcal{D}_{0}^{-1}\;D_{\pi}\;\mathcal{D}%
_{0}^{-1}\;D_{a}\right]  +\;\mathrm{Tr}\left[  \mathcal{D}_{0}^{-1}\;D_{\pi
a}\right]  \ , \label{spia}%
\end{equation}
where, $\mathcal{D}_{\pi}$, $\mathcal{D}_{a}$ and $\mathcal{D}_{\pi a}$
stand for the terms in the expansion of Eq.~(\ref{aa}) that are linear in
$\delta \pi_i$ and/or $a_{\mu}$. The corresponding expressions are long and
will not be quoted here. The $\pi$DA within the nlNJL model can then be
obtained by taking the functional derivative of $S_{\pi a}$ with respect
to $\delta\pi_i$ and $a_{\mu}$. It is important to note that due to the
bilocal character of the gauge field $a_{\mu}$ associated with the current
in Eq.~(3) an extra delta function appears in momentum space. Namely,
while for the local case we would have
\[
\int d^{4}x\,\bar{\psi}\left(  y\right)  \Gamma \psi\left(  y\right)
\,e^{iq\cdot y}=\int\frac{d^{4}p_{1}}{\left(  2\pi\right)  ^{4}}\frac{d^{4}p_{2}%
}{\left(  2\pi\right)  ^{4}}\left(  2\pi\right)  ^{4}\delta^{(4)}\left(
p_{2}+q-p_{1}\right)  \,\bar{\psi}_{p_{2}}\,\Gamma\,\psi_{p_{1}}\ \ \ ,
\]
for a bilocal current of the type appearing in Eq.(3) we have
\begin{align*}
&  \int\frac{d\xi^{-}}{2\pi}\int d^{4}x\left.  \,\bar{\psi}(y-\xi/2)
\,\Gamma\,\psi(y+\xi/2)\right\vert
_{\xi^{+}=0,\,\vec{\xi}_{T}=0}\,e^{iq\cdot y}\,e^{iP^{+}\xi^{-}x}\\
&  \qquad \qquad =\ \int\frac{d^{4}p_{1}}{\left(  2\pi\right)^{4}}
\frac{d^{4}p_{2}}{\left(
2\pi\right)  ^{4}}\left(  2\pi\right)  ^{4}\delta^{(4)}\left(  p_{2}%
+q-p_{1}\right)  \,\delta\left(  P^{+}x-\frac{p_{1}^{+}+p_{2}^{+}}{2}\right)
\,\bar{\psi}_{p_{2}}\,\Gamma\,\psi_{p_{1}}\ ,
\end{align*}
where $\Gamma$ represents an operator carrying Dirac and flavor indices.
In this way, besides the delta function related to four-momentum
conservation one has an extra one-dimensional delta that involves the $+$
components of the momenta. The latter can be worked out in Minkowski space
[e.g.\ by integrating over the $z$ component of the momentum $k^\mu \equiv
\frac12 (p_1+p_2)^\mu$], going then back to Euclidean space.

\begin{figure}
[ptb]
\begin{center}
\includegraphics[
height=1.3765in,
width=3.5483in
]%
{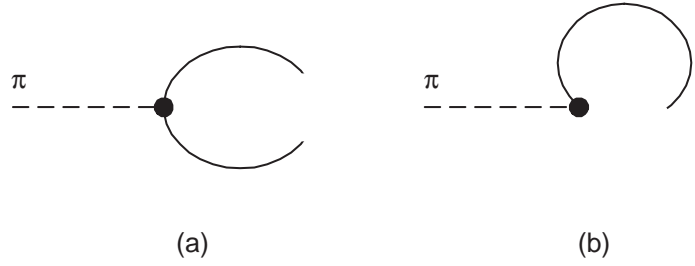}%
\caption{Diagrammatic representation of the contributions to the $\pi$DA.
As usually, in diagram (a) the struck quark connects the two open quark
lines, whereas in diagram (b) the struck quark goes from the open quark
line to the pion-quark vertex.}%
\label{Fig_Diagram}%
\end{center}
\end{figure}

The contributions coming from the two terms in Eq.~(\ref{spia}) can be
represented diagrammatically as shown in Fig.~\ref{Fig_Diagram}, where
Diag.~(a) corresponds to the first term and Diag.~(b) to the second one.
Regarding the expressions in Eqs.~(\ref{terms}-\ref{DA.11}), Diag.~(b) gives
rise to the last term of $F_{2}$ [see Eq.~(\ref{DA.11})] while Diag.~(a)
accounts for $F_{1}$ and the remaining terms in $F_{2}$.

\section{Renormalization factors for the QCD evolution of the $\pi$DA}

\label{App_QCD_Evolution}

We quote here the expressions for the renormalization factors $E_{n}%
^{\text{LO}}$, $E_{n}^{\text{NLO}}$ and $d_{n}^{k}$ needed to calculate
the evolution of the coefficients $a_{n}(\mu)$ in Eqs.~(\ref{DA.34}) and
(\ref{DA.35}). One has%
\begin{align}
E_{n}^{\text{LO}}(\mu,\mu_{0}) &  =\left[  \frac{\alpha_{\mathrm{s}}(\mu
)}{\alpha_{\mathrm{s}}(\mu_{0})}\right]  ^{\gamma_{n}^{(0)}/(2\beta_{0}%
)}\ ,\label{ap.b01}\\
E_{n}^{\text{NLO}}(\mu,\mu_{0}) &  =E_{n}^{\text{LO}}(\mu,\mu_{0})\left[
1+\frac{\alpha_{\mathrm{s}}(\mu)-\alpha_{\mathrm{s}}(\mu_{0})}{8\pi}%
\frac{\gamma_{n}^{(0)}}{\beta_{0}}\left(  \frac{\gamma_{n}^{(1)}}{\gamma
_{n}^{(0)}}-\frac{\beta_{1}}{\beta_{0}}\right)  \right]  ,\nonumber
\end{align}
where $\beta_{0}\,(\beta_{1})$ and $\gamma_{n}^{(0)}(\gamma_{n}^{(1)})$
are the LO (NLO) coefficients of the QCD $\beta$-function and the
anomalous dimensions, respectively. The first two coefficients of the
$\beta$-function are
\begin{equation}
\beta_{0}=11-\frac{2}{3}\,n_{f}\,,\qquad\beta_{1}=102-\frac{38}{3}%
\,n_{f}\ ,\label{ap.b02}%
\end{equation}
where $n_{f}$ is the number of flavors (we take here $n_{f}=4$). For the
evolution of the strong coupling constant $\alpha_{s}$ we use
\begin{equation}
\alpha_{s}(\mu)\ =\ \frac{4\pi}{\beta_{0}\ln(\mu^{2}/\Lambda^{2})}\left\{
1\;-\;\frac{\beta_{1}}{\beta_{0}^{2}}\;\frac{\ln\big[\ln(\mu^{2}/\Lambda
^{2})\big]}{\ln(\mu^{2}/\Lambda^{2})}\right\}  \ ,
\end{equation}
taking $\Lambda=0.224$~GeV ($\Lambda=0.326$~GeV) if the calculation is
carried out at the LO (NLO). The anomalous dimensions $\gamma_{n}^{(0)}$
are given by
\begin{equation}
\gamma_{n}^{\left(  0\right)  }=2\,C_{F}\left(  1-\frac{2}{(n+1)(n+2)}%
+4\sum_{m=2}^{n+1}\frac{1}{m}\right)  \ ,\label{ap.b03}%
\end{equation}
while analytical expressions for $\gamma_{n}^{\left(  1\right)  }$ can
be found in Refs.~\cite{Floratos:1977au,GonzalezArroyo:1979df}.

On the other hand, the off-diagonal mixing coefficients $d_{n}^{k}$ in
Eq.~(\ref{DA.35}) are given by:
\begin{equation}
d_{n}^{k}(\mu,\mu_{0})\ =\ \frac{M_{n}^{k}}{\gamma_{n}^{(0)}-\gamma_{k}%
^{(0)}-2\beta_{0}}\left\{  1-\left[  \frac{\alpha_{\mathrm{s}}(\mu)}%
{\alpha_{\mathrm{s}}(\mu_{0})}\right]  ^{[\gamma_{n}^{(0)}-\gamma_{k}%
^{(0)}-2\beta_{0}]/2\beta_{0}}\right\}  \ .\label{ap.b04}%
\end{equation}
Here the matrix $M_{n}^{k}$ is defined as
\begin{align}
M_{n}^{k} &  =\frac{(k+1)(k+2)(2n+3)}{(n+1)(n+2)}\left[  \gamma_{n}%
^{(0)}-\gamma_{k}^{(0)}\right]  \nonumber\\
&  \times\left\{  \frac{8C_{F}A_{n}^{k}-\gamma_{k}^{(0)}-2\beta_{0}%
}{(n-k)(n+k+3)}+4C_{F}\frac{A_{n}^{k}-S_{1}(n+1))}{(k+1)(k+2)}\right\}
\ ,\label{ap.b05}%
\end{align}
where
\begin{equation}
A_{n}^{k}=S_{1}\left(  \frac{n+k+2}{2}\right)  -S_{1}\left(  \frac{n-k-2}%
{2}\right)  +2\,S_{1}(n-k-1)-S_{1}(n+1)\ .\label{ap.b06}%
\end{equation}
Numerical values of the coefficients $M_{n}^{k}$ for $n\leq12$ are given
in Ref.~\cite{Agaev:2010aq}.

\end{document}